\documentclass{emulateapj} 

\newcommand{\icarus}{Icarus}
\begin{document}

\title{Resonance Trapping in Protoplanetary Disks. I. Coplanar Systems}

\author{Aaron T. Lee\altaffilmark{1,2}, Edward W. Thommes\altaffilmark{1}, and Frederic A. Rasio\altaffilmark{1}}
    
\email{alee@astro.berkeley.edu}

\altaffiltext{1}{Department of Physics and Astronomy, Northwestern University, Evanston, IL 60208}
\altaffiltext{2}{Department of Astronomy, University of California - Berkeley, Berkeley, CA 94720 }

\begin{abstract}
Mean-motion resonances (MMRs) are likely to play an important role both during and after the lifetime of a protostellar gas disk. We study the dynamical evolution and stability of planetary systems containing two giant planets on circular orbits near a 2:1 resonance and closer.  We find that by having the outer planet migrate inward, the two planets can capture into either the 2:1, 5:3, or 3:2 MMR. We use direct numerical integrations of $\sim 1000$ systems in which the planets are initially locked into one of these resonances and allowed to evolve for up to $\sim 10^7$~yr.  We find that the final eccentricity distribution in systems which ultimately become unstable gives a good fit to observed exoplanets.  Next, we integrate $\sim 500$ two-planet systems in which the outer planet is driven to continuously migrate inward, resonantly capturing the inner; the systems are evolved until either instability sets in or the planets reach the star.  We find that although the 5:3 resonance rapidly becomes unstable under migration, the 2:1 and 3:2 are very stable.  Thus the lack of observed exoplanets in resonances closer than 2:1, if it continues to hold up, may be a primordial signature of the planet formation process.
\end{abstract}

\section{Introduction}
The discovery of extrasolar planets around sun-like stars \citep{mayorqueloz95,marcybutler95,marcybutler98} has revealed that early large-scale migration likely plays an important role in the formation of planetary systems. For example, observations have revealed a nontrivial number (roughly 6\% overall) of close orbiting gas giants, called `hot Jupiters,' planets that would have great difficulty forming in their present locations \citep{bodenheimeretal00}. In fact, disk-planet interaction theory predicts disturbingly short migration time scales that seem to threaten the survival of everything ranging from planetary embryos to gas giants \citep{ward97}. Additionally, there are currently at least eight planetary systems that have two planets in MMR \citep{udryetal07}. These include systems such as GJ 876 \citep{marcyetal01} and HD 82943 \citep{leeetal06} with two planets in a 2:1 resonance, and HD12661 \citep{fischeretal03}, which contains two planets in what may be a 6:1 resonance. The origins of such resonant systems are thought to be due to disk-planet interactions that induce angular momentum transfer and differential migration  \citep{snellgroveetal01,leepeale02,papa03,kleyetal04}. 

There are currently two competing theories to explain planet formation. The ``core accretion" model starts with the sedimentation and collisional growth of dust grains and smaller planetesimals in the protoplanetary disk \citep[see][and references therein]{lissauer93}. These rocky cores (protoplanets) continue to build up until their gravitational influence allows for the accretion of the surrounding gas, forming gas giants. The other theory invokes direct gravitational instabilities in the disk \citep[e.g.,][]{cameron78,boss00}. 

In these theories the dynamical relaxation time of a planetary system can, in principle, be longer than the time scale for planet formation. Therefore, long term stability is not guaranteed for a planetary system's initial configuration. Gravitational interactions between massive protoplanets can lead to orbit crossing and instability, resulting in massive planets being thrown closer to their parent stars \citep{rasioford96,weidenschillingmarzari96}. In the case of one of the planets being ejected, the surviving planet is left with high eccentricity (some with $e > 0.90$). Additionally, migration occurs within the disk due to the exchange of angular momentum between the planet and the surrounding disk material through planet-disk interactions \citep{goldtrem79,goldtrem80,linpapa79,linpapa93,papalin84,ward86}. In a laminar disk, there are two basic modes of migration: For an embedded body, imbalance between the torques from the inner and outer parts of the disk is thought to lead to orbital decay; this is commonly referred to as type I migration \citep[e.g.,][]{ward97}.  If a body is massive enough, it locks itself to the disk by opening a deep annular gap, and is thus carried along as the disk accretes onto the star \citep{linpapa86,ward97}; this is called type II migration.\footnote{There is also a proposed type III migration, which involves disk material flowing through the co-orbital region of the planet, which will not be considered here \citep[see][]{papaetal07}.} When orbital migration occurs for different planets at different rates, convergent migration and locking into mean-motion commensurabilities can occur. 

For the two planet case, the regions of dynamical stability (i.e., the region where close encounters are impossible) can be determined analytically by studying how the value of the Jacobi integral relates to the topological (Hill) stability of the system \citep{gladman93}. However, this criterion does not take into account the fact that planets may be in resonance. It has been shown that resonant systems have additional regions of stability \citep{barnesgreen07}, which current analytic criteria predict to be unstable. In fact, \cite{barnesgreen07} have shown that nearly all observed resonant systems lie in these extended regions.

Overall, planet-planet-disk interactions are likely key in determining the final configuration of a planetary system. To date, each type of dynamical process (planet-disk interactions, dynamical instabilities, and resonances) has usually been considered individually. The motivation of this study is to explore physical situations where all of these processes occur simultaneously. In particular, we consider gap-opening planets located close to particular resonances within a protoplanetary disk. This results in convergent migration and resonance capture, but possibly also eventual instability due to close encounters. 

There are two objectives to this study. The first is to better quantify the regions of stability for particular resonances. More explicitly, we will consider the 3:2, 5:3, and 2:1 resonance for two gap-opening planets of varying mass. We place an upper bound on the possible masses for planets in these three resonances. This provides an interesting problem in dynamics by allowing us to more generally map the parameter space for these particular resonances in and out of the protoplanetary disk, compared to \citet{barnesgreen07} who studied particular observed extrasolar systems. Additionally, in the case where there are instabilities, this study allows us to see how the final configurations (e.g., distribution of eccentricity) differ, if at all, from non-resonant planet scattering \citep[see, e.g.,][]{fordetal01,chatterjeeetal07}. 

Our second objective is to study the dynamics of gas giants in the region of the protoplanetary disk where we believe these planets form, between 5 and 10 AU. The three above resonances are typical resonances that planets can resonantly capture into, since they require little to no eccentricity to give a high probability of capture when planets migrate at rates consistent with Type II migration. Here we are assuming that planets are born on nearly circular orbits. We then study the evolution as the coupled system of two planets migrates with the disk. To date, the closest resonance to be observed---that is, the resonance with the smallest ratio of outer to inner period---is the 2:1. We aim to better quantify why we have not found any systems in the 3:2 or 5:3 resonances and measure how common we should expect each of these resonances to be in extrasolar planetary systems.

This paper is organized as follows. In \S 2, we explain the numerical treatment used in both the resonance stability and disk simulations. In \S 3, we give further details of the theory used in our models for studying the stability of the aforementioned resonances. We then present the results of those simulations. In \S 4, we develop the theory used to model the planet-disk interactions in the protoplanetary disk as well as the results of those runs. We conclude with a summary and discussion in \S 5.  

\section{Numerical Methods}
Newton's equations of motion are integrated directly in the barycentric frame using a fifth-order Runge-Kutta (R-K) scheme \citep{press07}. Although these direct integrations are computationally more taxing compared to symplectic integrations, they are able to handle arbitarily short dynamical  times (e.g., when the planets are orbiting within a few solar radii of the star). Since our integrations run at most a length of $10^7$ yr, we allow the R-K scheme to incur a maximum error of order 1 part in $10^9$ per time step, where the time step in the three-body problem is variable. This ensures that the total accumulated error in energy, characterized through the ratio $\Delta\,E/E_i$, where $E_i$ is the initial system energy and $\Delta\,E$ is the difference between this value and the final energy of the system, remains less than $10^{-6}$ in the conservative case. Similarly, $\Delta\,A/A_i$ remains less than $5\cdot 10^{-6}$, where $A_i$ is the system's initial angular momentum, ensuring that the accumulated errors do not affect the final result.

In all our simulations, we consider a two-dimensional three-body system, consisting of a solar-mass star and two planets with masses ranging between 0.25 Jupiter-masses ($M_J$) to 12 $M_J$. The system evolves until either a planet is ejected, a collision between any two bodies occurs, or the system evolves for $10^7$ yr. An ejection is defined as a planet being at least 250 AU from the star with positive energy. A collision between the star and a planet occurs when the relative distance between the two is less than $1.1\ R_\sun$, where $R_\sun$ is the radius of a solar-mass star.\footnote{$1\ R_\sun+1\ R_J\approxeq 1.1\ R_\sun$, where $R_J$ is the radius of Jupiter.} Similarly, a collision between two planets occurs when the relative distance is less than $2\ R_J$. The simulation continues for several years after the collision or ejection before ending the run. In the event of a collision, the pair is replaced with a single point mass with the same linear momentum as the original pair and a mass as simply the sum of the original masses. 

In these simulations, the units are selected so that $1$ AU $=1$, $M_\sun=1$, and $G=1$, where $G$ is the gravitational constant. The planet's initial orbits are always set to be circular. 

\section{Resonance Stability}
\label{sec:resonance stability}

As mentioned in Section 1, the region of dynamical stability is sharply defined for the two planet case, and its location is analytically known \citep{gladman93}. That is, given the masses of the two planets, the minimum orbital separation in the semi-major axes to prevent close encounters can be determined. Since we wish to consider two planets in a particular resonance, we can reverse the criterion to give analytically determined limits on the masses of the planets. We compute the separation in semi-major axes via Kepler's third law. Since the ratio of the orbital periods $T_o/T_i>1$, where the subscripts denote the outer and inner planet, respectively, we get the following relation: $a_o=(T_o/T_i)^{2/3}a_i$. Therefore, the orbital separation normalized by the inner planet's semi-major axis is simply
\begin{eqnarray}
	\Delta&\equiv&\frac{a_o-a_i}{a_i}=\frac{p^{2/3}a_i-a_i}{a_i}=p^{2/3}-1,
\end{eqnarray}
where we have defined $p=T_o/T_i$. So $\Delta$ is a constant for a given resonance. Now the analytic criterion for stability for low-eccentricity planets is given by 
\begin{eqnarray}
\label{eqn:glad}
	\Delta_{min}&\equiv&2\cdot 3^{1/6}(\mu_1+\mu_2)^{1/3}+2\cdot 3^{1/3}(\mu_1+\mu_2)^{2/3}\nonumber \\
	&&-\frac{11\mu_1+7\mu_2}{3^{11/6}(\mu_1+\mu_2)^{1/3}}+\cdots,
\end{eqnarray}
where $\mu_1$ ($=M_1/M_\sun$) is chosen to be the larger of the two bodies \citep{gladman93}. By setting $\Delta=\Delta_{min}$, we can derive a maximum limit on the masses of the planets to ensure stability, as suggested by Equation \ref{eqn:glad}. Table 1 gives some results for equal mass planets. Additionally, it compares the results whether one uses only one, two or three terms in the expansion (Equation \ref{eqn:glad}). One can see for resonances with larger orbital separations, truncating the series with the first term is inadequate to accurately describe the analytical limit. In fact, the addition of the second term results with over 100\% correction to the mass limit for resonances above the 2:1 MMR.

\begin{deluxetable*}{|c|c|c|c|c|c|}
\tablecaption{Mass Limits for particular resonances as determined by Equation \ref{eqn:glad}. Here $p$ denotes the ratio of the outer and inner planet's period, $T_o/T_i$. Limits are given depending on whether only one (first-order), two, or all three terms are used in Equation \ref{eqn:glad}.}
\tablewidth{0pt}
\startdata \hline
Resonance & $p$ & $M_1+M_2$ & $M_1+M_2$ & $M_1+M_2$ & First-Order /\\ 
& & ($M_J$, first-order) & ($M_J$, second-order) & ($M_J$, third order) & Third-Order \\ \hline
3:2 & 1.50 & 2.26 & 1.54 & 1.77 & 1.272\\
5:3 & 1.66 & 5.05 & 3.13 & 3.72 & 1.356\\
2:1 & 2.00 & 15.33 & 8.09 & 10.10 & 1.517\\
5:2 & 2.50 & 45.15 & 19.66 & 25.86 & 1.745\\
3:1 & 3.00 & 95.29 & 35.57 & 48.58 & 1.961\\ 
\hline

\enddata
\end{deluxetable*}

\subsection{Initial Conditions}
\label{sec:initial conditions}
We randomize the initial semi-major axis of the inner planet between 5 and 6 AU. For the 3:2 resonance, the masses of both planets are randomized between 0.25 $M_J$ and 4 $M_J$. Similarly, for the 5:3 resonance, the masses are randomized between 0.25 $M_J$ and 5 $M_J$, and the 2:1 between 1 $M_J$ and 12 $M_J$. The initial phase of the inner planet is set to zero, while the outer planet's is randomized between 0 and $2\pi$.  

To determine the initial position of the outer planet, we set the planet's semi-major axis so that it is just outside the particular resonance by a small fixed amount ($\approx 0.1$ AU). We then apply a frictional force of the form $-\alpha\mathbf{v}$, where $\alpha$ is a constant and $\mathbf{v}$ is the velocity of the outer planet. It is applied for a short amount of time to allow the outer planet to migrate into the resonance location with the inner planet. The time and strength have been adjusted so no significant coupled migration occurs. Once the planets begin to migrate together in resonance, the change in their semi-major axis becomes correlated with the change in their eccentricity \citep[see][]{murraydermott}. Since no significant coupled migration occurs, the eccentricity of each planet never grows above values of 0.2 from migration/eccentricity excitation alone. 

To monitor the resonance in question, we use the resonance variable defined by
	\begin{equation}
	\varphi=j_1\lambda_o+j_2\lambda_i+j_3\omega_o+j_4\omega_i
	\label{eqn:resvar}
	\end{equation}
\citep{murraydermott}. Here $\lambda$ and $\omega$ are the mean longitude and longitude of pericenter, respectively. The subscripts denote either the inner or outer planet. Since our integrations are two-dimensional, the terms including the longitude of the ascending node ($\Omega$) are zero. For a $p:q$ resonance, the ordered set $\{j_1,j_2,j_3,j_4\}$ is equal to $\{p,-q,-(p-q),0\}$ or $\{p,-q,0,-(p-q)\}$. When two planets are in a given resonance with pericenter alignment, the resonance variable is bounded with a libration amplitude $< 2\pi$. Additionally, planets can still be locked in resonance although $\varphi$ continues to oscillate between 0 and $2\pi$. Such configurations require oscillations of both the semi-major axis and eccentricity to account for the circulating longitude of pericenters \citep{murraydermott}. 

A run is defined as stable if the system evolves for $10^7$ yr and maintains the original configuration of semi-major axes without appreciable changes in any orbital elements (eccentricity, etc.). In these runs, the planets must lock into a mean motion resonance so that $\phi$ becomes bounded with a finite libration amplitude or circulates in a periodic manner. A run is defined as unstable if the system goes unstable before or after the planets lock into mean motion resonance.
\subsection{Results}
We present the results for the resonance stability tests, analyzing both the region of stability for these particular resonances as well as the final configuration of the systems that went unstable. We first focus on each resonance separately, and will consider broader results in the final section.

	\subsubsection{The 2:1 Resonance}
We determined analytically from Equation \ref{eqn:glad} that the 2:1 resonance is stable up to $M_i+M_j \approx 10\ M_J$. Figure \ref{fig:21Stab} presents a stability map, where the filled circles represent a stable run, and the (red) open circles represent runs where instability occurred before the planets could lock into a 2:1 MMR. The dashed line in the figure is the analytical boundary between stable and unstable regions determined by Equation \ref{eqn:glad}. It includes all three terms and hence is not symmetric in the masses. It's slope therefore changes depending on whether the inner or outer planet is the more massive one (recall the definition of $\mu_1$ in Equation \ref{eqn:glad}). The upper left region of the plot shows an obvious region of stability not predicted by the simple analytic criterion. That is, the additional stability typically involves a larger outer planet, and includes up to two $11\ M_J$ planets remaining stable. However, there is another obvious region of instability not predicted by Equation \ref{eqn:glad} for an inner planet roughly twice the mass of the outer planet. In these configurations, the planets went unstable due to the inner planet exciting the eccentricity of the outer planet. This resulted in the outer planet undergoing a close encounter with the inner planet and typically was ejected from the system. 

Over 90\% of the unstable runs resulted in the less massive planet being ejected from the system, leaving the remaining planet with eccentricities that typically range from 0.3 all the way to 0.97. The latter of these would be expected to undergo strong tidal forces at periastron and result in a less eccentric orbit with a smaller semi-major axis (see Section \ref{sec:ppdinteract}). The time scales to undergo instability are typically very short. Most of the runs that go unstable before entering resonance do so within $10^4$ yr. Once the planets entered resonance to a point where $\phi$ become bound and pericenter alignment occurred, the planets enjoyed a phase protection that prevented any further close encounters.

	\subsubsection{The 3:2 Resonance}
	
\begin{figure}
\plotone{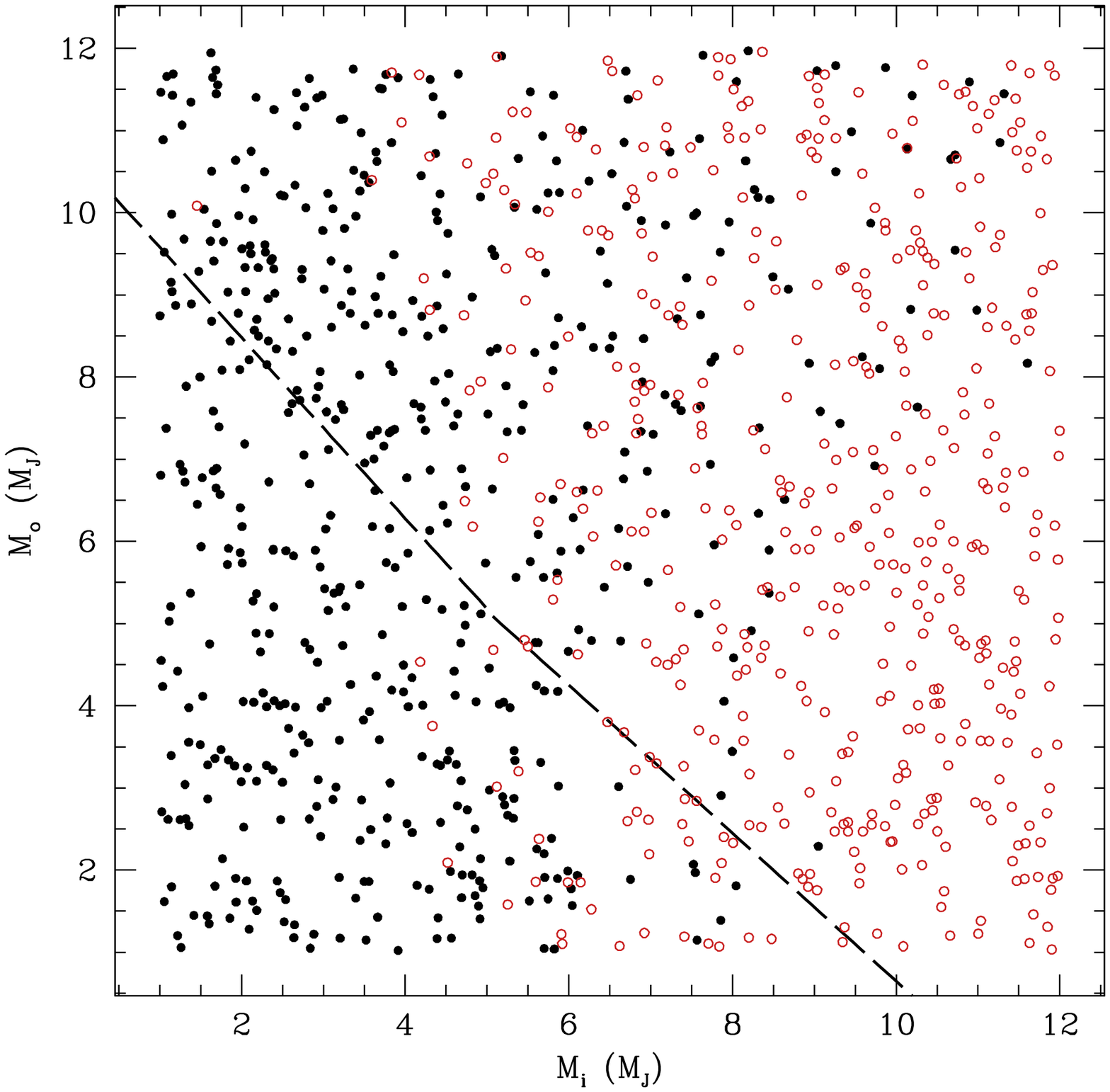}
\caption{Stability map for planets initially on circular orbits in the 2:1 resonance. A filed circle shows a run where the planets remained in the resonance for $10^7$ yr, and an open (red) circle \ shows runs that went unstable before entering a 2:1 MMR. Planets were determined whether to be in resonance by monitoring the resonance variable defined by Equation \ref{eqn:resvar}. The analytical boundary between stability and instability determined by Equation \ref{eqn:glad} is shown by the dashed line, with the analytical region of stable lying below the line, and unstable otherwise.} 
\label{fig:21Stab}
\end{figure}

\begin{figure}
\plotone{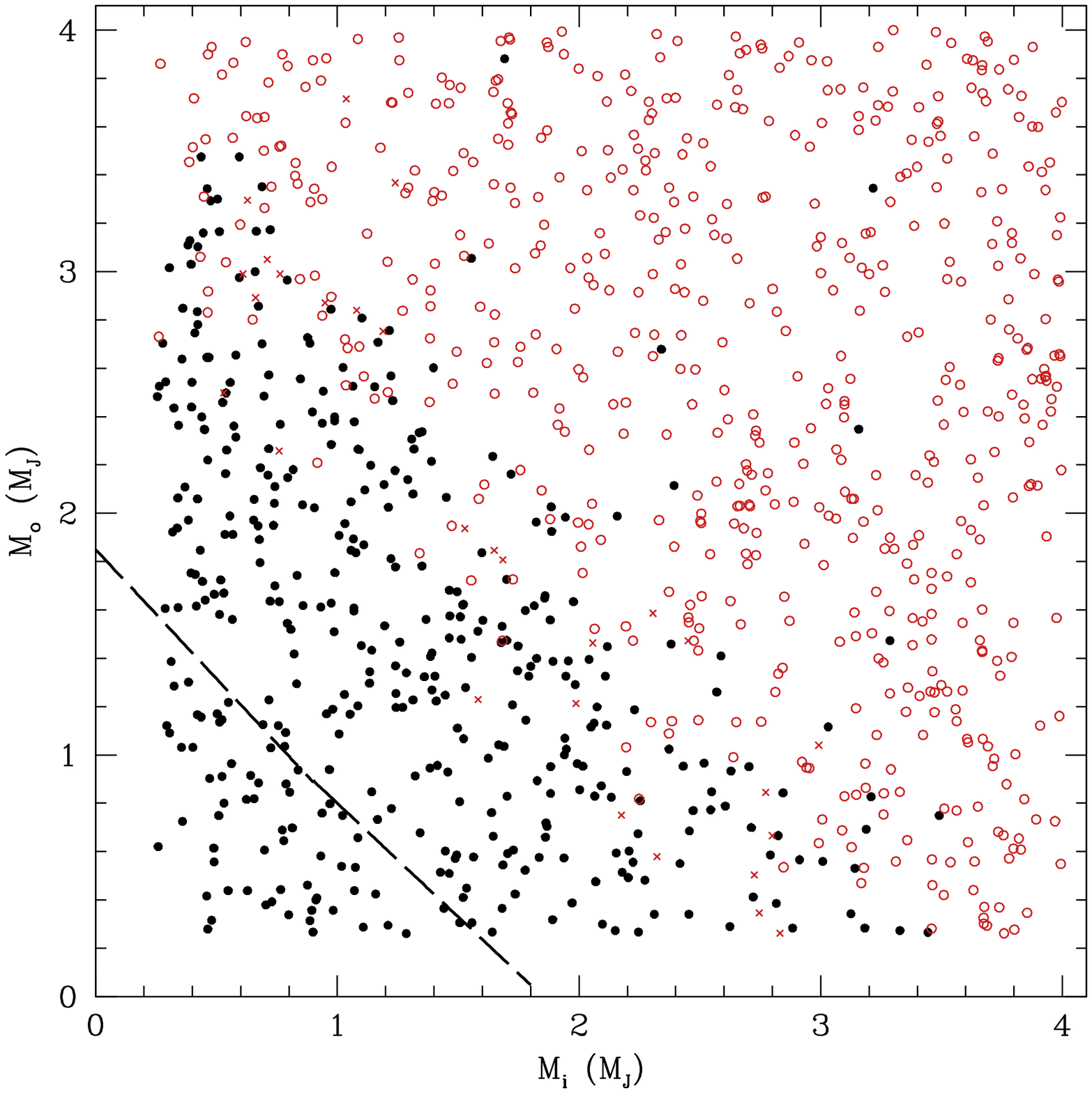}
\caption{Stability map for planets initially on circular orbits in the 3:2 resonance. A filed circle shows a  run where the planets remained in the resonance for $10^7$ yr, and an open (red) circle or cross shows runs that went unstable before or after entering a 3:2 MMR, respectively. Planets were determined whether to be in resonance by monitoring the resonance variable defined by Equation \ref{eqn:resvar}. The analytical boundary between stability and instability determined by Equation \ref{eqn:glad} is shown by the dashed line, with the analytical region of stable lying below the line, and unstable otherwise.} 
\label{fig:32Stab}
\end{figure}

Equation \ref{eqn:glad} suggests that the 3:2 resonance is stable up to $M_i+M_j \approx 2\ M_J$. This region of stability is plotted in Figure \ref{fig:32Stab}, which uses the same conventions as Figure \ref{fig:21Stab}. Additionally, the (red) crosses represent runs where the planets showed signs of being in a 3:2 MMR but eventually went unstable. The region of stability for the 3:2 resonance is more symmetric compared to the 2:1 resonance, with the region of stability extending to over twice the analytically determined values. The 3:2 resonance typically does not include any configurations where one of the planets is more than 3 $M_J$. 

The majority of the instabilities result in a collision, leaving the remaining planet in a position between the initial semimajor axes of the two original planets with a very low ($<0.1$) eccentricity. In the several cases where the instability did not immediately result in a collision and the two planets strongly interacted for an extended period of time, the simulation resulted in either a collision or ejection, both leaving the remaining planet with a higher final eccentricity. A low number of runs ($<5\%$) resulted in the planets being thrown into a configuration that did not go unstable for the remainder of the of the integration. In these cases, the ratio of periods, $T_2/T_1$ remained in the interval $\left[5.0 , 9.0\right]$.

\begin{figure}
\plotone{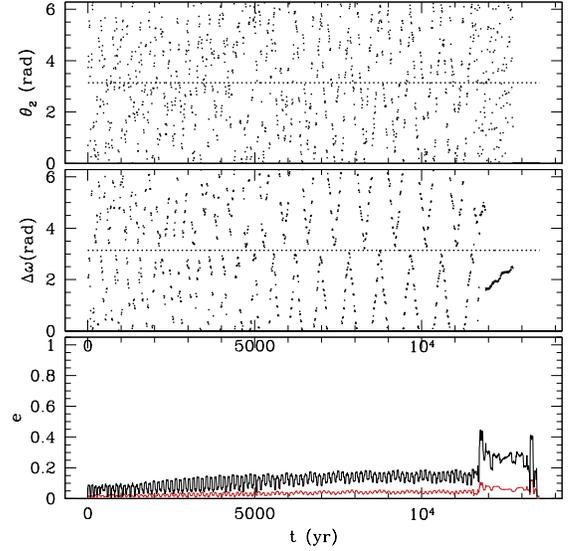}
\caption{Evolution of two planets that enter a 3:2 MMR but eventually become unstable and result in a collision. The top panel plots one of the resonance variables, defined by Equation \ref{eqn:resvar}, while the middle panel plots the pericenter difference and the bottom panel plots the eccentricity of the inner and outer planet (black and red lines, respectively). The two planets enter resonance around 5000 yr, but strong gravitational perturbations break the resonance around $1.1\cdot 10^4$ yr.} 
\label{fig: 32resunstable}
\end{figure}

There exists a few cases where planets begin to enter a 3:2 MMR but eventually go unstable. These instabilities occur in less than $10^5$ yr.  An example is shown in Figure \ref{fig: 32resunstable}. The top panel plots one of the resonance variables:
	\begin{equation}
	\varphi=p\lambda_o-q\lambda_i-(p-q)\omega_o,
	\end{equation}
where the middle panel plots difference in pericenters of the two planets, and the bottom panel plots the eccentricity of the inner and outer planet (black and red lines, respectively). We see that the planets show sign of entering resonance around 5000 yr, where both $\varphi$ and $\Delta\omega$ show periodic variation from that point onward. However, shortly after $1.1\cdot10^4$ yr, the two planets leave the resonance and go unstable due to strong gravitational perturbations. This particular run resulted in a collision. 

	\subsubsection{The 5:3 Resonance}

\begin{figure}
\plotone{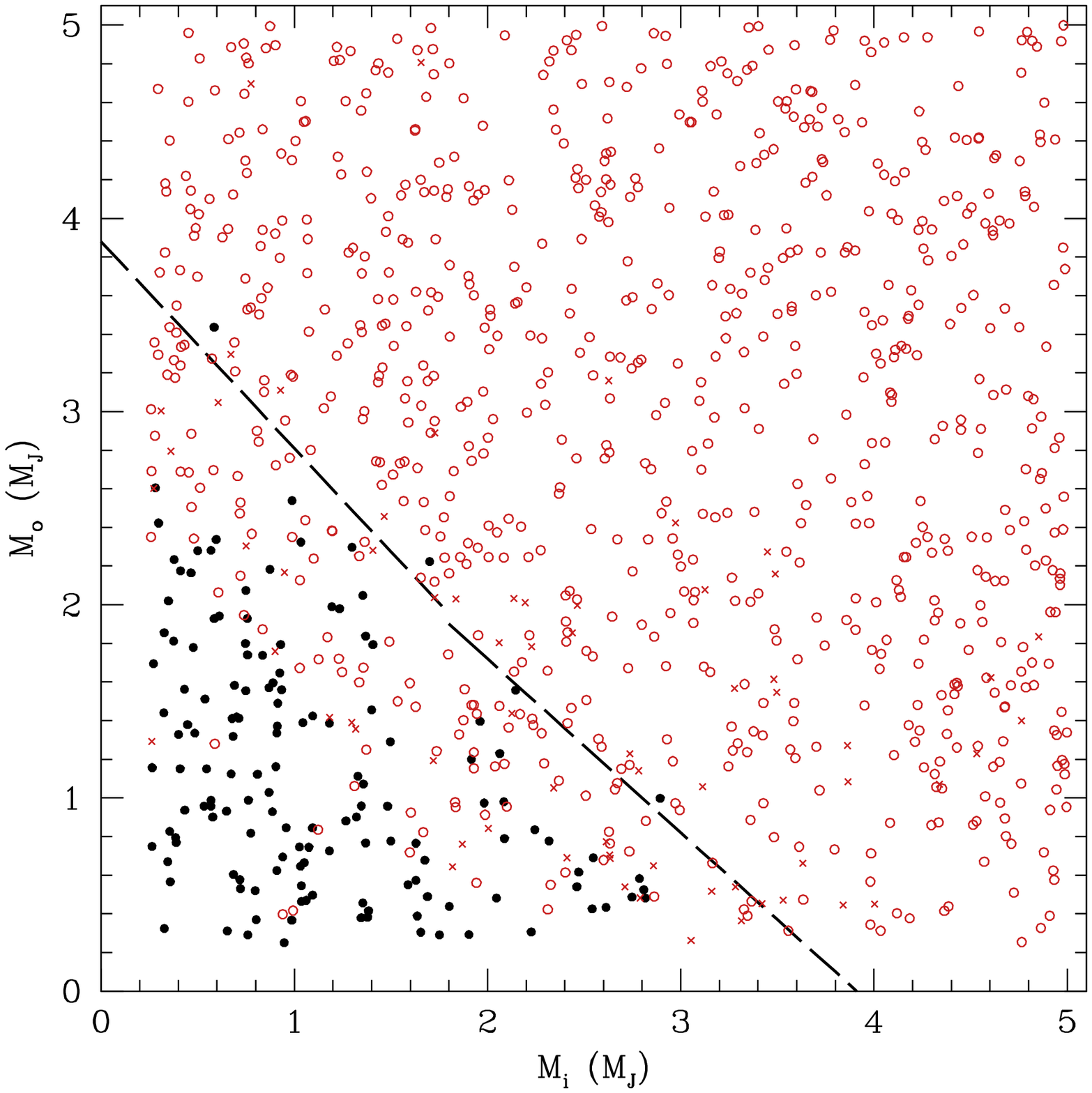}
\caption{Stability map for planets initially on circular orbits in the 5:3 resonance. A filed circle shows a  run where the planets remained in the resonance for $10^7$ yr, and an open (red) circle or cross shows runs that went unstable before or after entering a 5:3 MMR, respectively. Planets were determined whether to be in resonance by monitoring the resonance variable defined by Equation \ref{eqn:resvar}. The analytical boundary between stability and instability determined by Equation \ref{eqn:glad} is shown by the dashed line, with the analytical region of stable lying below the line, and unstable otherwise.} 
\label{fig:53Stab}
\end{figure}

Finally, Figure \ref{fig:53Stab} shows the stability map for the second-order resonance\footnote{All resonances can be written in the form $(a+b)/a$, where $b$ denotes the order of the resonance.}, the 5:3. Compared to the 2:1 and 3:2 MMRs, the region of stability is less than that predicted by analytic criteria with only three stable runs existing beyond the region of analytically determined stability. These three runs, however, are quite close to the border between the two analytically determined regions. Additionally, a number of unstable runs exist deep within the region predicted to be stable, showing that configurations in a 5:3 resonance are very sensitive to their initial conditions. The majority of the simulations result in an ejection, leaving the final planet with an eccentricity in the interval $[0.3,0.8]$. Collisions result in outcomes similar to that of the 3:2 resonance: a merger product with low eccentricity somewhere between the semimajor axes of the two original planets.

\begin{figure}
\plotone{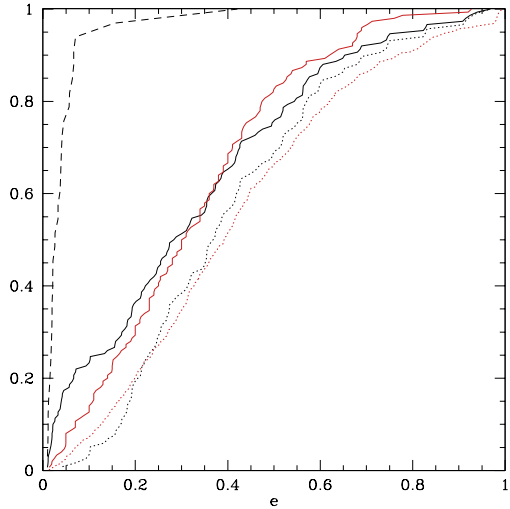}
\caption{Cumulative distribution showing the final eccentricity of the remaining planet after instability. It includes the data from all three resonances (the 3:2, 5:3, and 2:1) without the presence of disk forces. Those due to collisions and ejections are shown by the dotted black and dashed black lines, respectively. The solid red line is the eccentricity distribution of observed extrasolar planets. It omits systems which are believed to be in resonance or are close enough to their parent star to undergo tidal interactions. The solid black line is a randomly selected combination of 50 outcomes from each of the three resonances. This is used to graph the overall distribution from all possible outcomes. It is compared with the final eccentricity distribution due to non-resonant planet-planet scattering from \citet{chatterjeeetal07}, shown by the dotted red line.} 
\label{fig:ResCumHisto}
\end{figure}

The above three resonances are potentially very important for any pair of planets that form in reasonably close proximity to each other. Planets can be captured into these resonances without having appreciable initial eccentricity, making them highly likely candidates for planets within the protoplanetary disk, where disk effects are believed to initially keep planets on circular orbits. These resonance impose a strict geometrical configuration on the orbits of the planets in and near these resonances. These periodic interactions between the two planets can lead to instabilities, resulting in ejections or collisions. Figure \ref{fig:ResCumHisto} plots the cumulative distribution of the remaining planet's eccentricity after an ejection or collision occurred due to an instability that formed during the integration.

Figure \ref{fig:ResCumHisto} compares the eccentricity distribution with that of observed extrasolar planets (solid red line). Ejections (dotted black line) produce a similar shape distribution, but underproduce lower eccentricity planets. Collisions (dashed black line) produce the majority of planets with eccentricities below 0.1. In order to compare these various processes with observed eccentricities, we randomly select 50 outcomes (which can be both collisions or ejections) from each of the three resonances. We then plot that distribution (solid black line) in Figure \ref{fig:ResCumHisto}, and find it matches observed eccentricities better than collisions or ejections alone. For further comparison, we include the results of non-resonant planet-planet scattering (red dotted line) from \citet{chatterjeeetal07}.
	
\clearpage

\section{Planet-Planet-Disk Interactions}
\label{sec:ppdinteract}

In this section we model planets interacting with a protoplanetary disk. The previous section considered the stability of three resonances without the effects of planet-disk interactions. We now study the stability of these resonances while the planets undergo migration due to angular momentum exchange between the planet and the surrounding disk material. Doing so will incite further eccentricity growth for the two planets, which can greatly affect the stability of the system.

We model the protoplanetary disk as a pure gas disk with an inner hole that extends all the way to the central star. A self-consistent treatment of planet-disk interaction would require us to also model the feedback of the planet on the disk, which in turn would necessitate incorporating hydrodynamics into our simulation.  To simplify matters, we impose the ``closing in" of the inner hole independently of the two planets' positions, but at a rate consistent with results from theory \citep{ward97} and full hydrodynamic simulations \citep{ogilvielubow03}. Therefore, the disc extends only to the location of the outer planet, and the inner planet essentially sits in a ``gap," and does not undergo migration unless it is due to resonance interactions. Hydrodynamic simulations \citep{thommesetal08} have shown that systems with multiple gap-opening planets tend to quickly remove all of the material inside the outermost planet, which puts the configuration in exactly the state we have assumed. 

The ``radius" of the disk's inner edge is set to migrate linearly inward at a rate of 1 AU every $10^5$ yr. A Fermi function is then used to model the edge of the disk, allowing the friction effects to smoothly fade away. That is, given the ``radius" of the disk edge, $r_d$, the acceleration due to friction applied to each planet is modeled by the following equation:
\begin{equation}
\label{eqn:fric}
	\mathbf{a_f}=-F(a,r_d)\cdot\left(\frac{\mathbf{v}}{t_m}+2\frac{(\mathbf{v\cdot r})\mathbf{r}}{r^2t_e}\right),
\end{equation}
where $F(a,r_d)$ is the Fermi function
\begin{equation}
	F(a,r_d)=1-\frac{1}{1+\mbox{exp}\left[(a-r_d)/(0.05r_d)\right]}
\end{equation}
for a given semi-major axis $a$. The ``width" of the Fermi function ($0.05r_d$) is selected so only the outer planet feels non-negligible force from the protoplanetary disk unless the two planets are going to unavoidably collide.\footnote{More specifically, it is chosen so the inner planet is always four ``widths" away from the radius of the disk. It is easy to show that, at this distance, the inner planet will always feel less than 2\% of the full magnitude of the frictional effects.} As the inner disk edge contracts, the planet adjusts its position relative to the disk edge until it reaches an equilibrium location.  At this location, the torque it experiences is equal to that needed to make it migrate inward at the same rate as the disk edge. 

The two terms in Equation \ref{eqn:fric} model the migration and eccentricity damping, respectively. The migration time scale, $t_m$, is set to 6000 yr, consistent with the type I rate for a body with mass of order $10^2\ M_{\oplus}$ \citep{ward97}; the exact value is unimportant as long as it is short compared to the time scale of the disk edge contraction. The eccentricity damping time scale, $t_e$, is set equal to either a tenth the value of $t_m$, as suggested by previous analytical and numerical studies of eccentricity damping in disks \citep{goldtrem80, trillingetal98,nelsonetal00,papaetal01} or $10^{10}$ yr, so to model the disk without eccentricity damping. 

As planets locked in resonance migrate together, their change in eccentricity is related to the mass ratio and rate of change in the semi-major axis of the planets \citep{murraydermott}. For planets that migrate to less than a tenth of their original semi-major axis, appreciable eccentricity growth occurs. Therefore, our simulations take into account energy lost due to the close-in tidal interactions between eccentric planets and the central star. As a planet approaches from large distances with large eccentricities, the relative distance is short at periastron compared to the rest of the orbit. During this brief period, angular momentum and energy is exchanged between the star and planet. We adopt the approximation developed by \citet{papaterq01}, where the tidal interaction acceleration is written as
\begin{equation}
	\mathbf{a_{Tide}}=-\frac{3Gm_pR^5_*}{2\sqrt{\pi}jr^{11}}\cdot\frac{0.6R^3_p}{1+(R_p/R_*)^3}\cdot(r^2\mathbf{v}-(\mathbf{r\cdot v})\mathbf{r}),
\label{eqn:tides}
\end{equation}
where $R_*$ is the stellar radius and $R_p$ ($= j^2/2GM_*$) is the distance of closest approach for a parabolic orbit with angular momentum $j$. This model is an approximation of a small perturbation limit for a non-rotating star and the planet being on a parabolic (or highly eccentric) orbit. It neglects the effects of tides acting on the planet itself. Since these interactions include a term proportional to the velocity of the planet, changes in the semi-major axis occurs rapidly and the planet ``falls" into the central star on short time scales. Therefore, we do not include relativistic effects in our simulations since the planets do not spend considerable time at distances were these effects would be appreciable. 

\subsection{Initial Conditions}
	The initial semi-major axis of the inner planet is randomized between 5 and 6 AU, and the mass of both planets is fixed at 1 $M_J$. The outer planet's semi-major axis is then randomized between the minimal distance at which the planets are stable according to Equation \ref{eqn:glad} (which is very near at the 3:2 resonance for two $M_J$ planets) and slightly outside the 2:1 resonance, that is,
\begin{equation}
a_i+2.4\,(\mu_i+\mu_o)^{1/3}a_i\leq a_o \leq 2^{2/3}a_i+\varepsilon,
\end{equation}
where $\varepsilon$ is roughly $0.1\,a_i$ and $\mu$ is defined as in Equation \ref{eqn:glad}. The initial location of the ``radius'' of the disk edge is chosen so that the outer planet lies approximately four ``widths'' away from the radius, which allows the outer planet to travel along the outer edge of the disk as it migrates inward without any sudden changes in the external disk forces. 

Two separate sets ($N\approx 500$) of integrations are done, one with $t_e=0.1 t_m$ and the other with $t_e=10^{10}$ yr (hereafter referred to as $t_e=\infty$). 

\subsection{Results}

As is to be expected given the initial conditions described above---particularly the low initial eccentricities---all planets are captured into one of three resonances: 2:1, 5:3, and 3:2.  An important parameter is the amount of eccentricity damping due to the disk, which affects the overall stability of the planets in these resonances. We find that the 2:1 resonance is the most stable; all planets captured into it remain in the 2:1 until they have migrated close enough to the star for tidal interactions to pull the planets apart (dragging the inner planet into the star). The 3:2 is similar for the case where damping is present and eccentricities are kept small. When eccentricity damping is not present, the 3:2 resonance goes unstable at eccentricities larger than 0.3. The 5:3 resonance goes unstable for very low eccentricities and produces the largest variety of possible outcomes with and without eccentricity damping. 

Each run has four possible results. The planets can either (1) capture into a particular resonance and migrate together until the inner planet is close enough to tidally interact with the star and the resonance is broken, resulting in the inner planet colliding with the star. In this case, we classify the run as ``Stable;" (2) capture into a particular resonance and undergo instability, so that the two planets collide with each other. The few cases ($\approx 1\%$) where instability results in one of the planets colliding with the star are also included here; (3) undergo instability once in a resonance and result in one of the planets being ejected; (4) after migrating in a particular resonance, undergo instability but ultimately end up in a different resonance. This occurs because one planet is thrown into the disc and is driven back inward towards the disc edge, recapturing the inner planet into a new resonance. Once in this resonance, the planets either quickly become unstable or remain in the resonance for remainder of the integration. The cases that remain in this new resonance are labeled ``Scattered and Recaptured." If they quickly become unstable and result in a collision or ejection, they are classified in the appropriate category (2) or (3) above.

We examine each case separately, with and without damping. The results are given in Table~ \ref{tab:resoutcome}. 

	\subsubsection{Overall Results with $t_e/t_m=0.1$}
	
\begin{figure}
\plotone{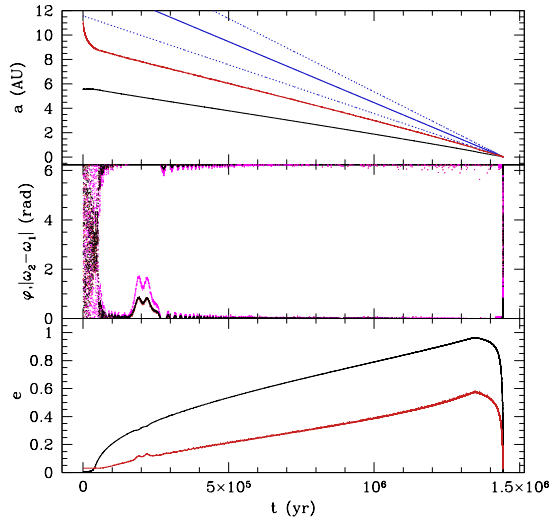}
\caption{The dynamical evolution of two planets in a 2:1 resonance with the presence of disk and tidal forces, letting $t_e/t_m=0.1$. \textit{Top Panel:} Plot of the semi-major axes of the outer and inner planet, shown by the red and black lines, respectively. The blue lines show the location of the disk's inner edge, with the outer blue lines being the width of the edge, and the middle line being the disk's inner radius. \textit{Middle Panel:} A plot of the absolute value of the difference in longitude of pericenters (black), and the two 2:1 resonance variables defined by Equation \ref{eqn:resvar} (red and magenta lines). The resonance breaks apart at the very end due to the presence of tidal forces acting on the inner planet, which pull the planets apart. \textit{Bottom Panel:} The eccentricity of the inner and outer planet, using the same color convention as the top panel. The decrease in eccentricities near the end of the evolution is due to the presence of tidal forces from the star, modeled by Equation \ref{eqn:tides}. } 
\label{fig:21Stable}
\end{figure}

\begin{figure}
\plotone{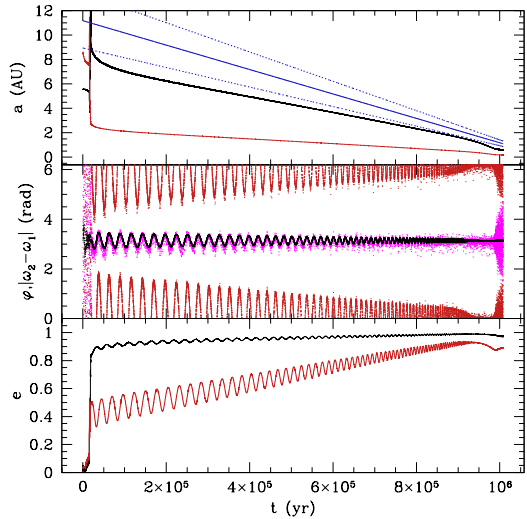}
\caption{The dynamical evolution of two planets that go unstable in a 5:3 resonance and get trapped into a 6:1 resonance. This occurs with the presence of disk and tidal forces, assuming $t_e/t_m=0.1$. \textit{Top Panel:} A plot of the semi-major axes of the outer and inner planet, shown by the red and black lines, respectively. The blue lines plot the location of the disk's inner edge, with the outer blue lines being the width of the edge, and the middle line being the disk's inner radius. \textit{Middle Panel:} A plot of the absolute value of the difference in longitude of pericenters (black), and the two 6:1 resonance variables defined by Equation \ref{eqn:resvar} (red and magenta lines). The resonance breaks apart at the very end due to the presence of tidal forces acting on the inner planet, which pull the planets out of the resonance. \textit{Bottom Panel:} The eccentricity of the inner and outer planet, using the same color convention as the top panel. } 
\label{fig:53Toss}
\end{figure}

We first present the case where eccentricity damping is present, which is set to be reasonably representative of planet-disk interactions (see Section \ref{sec:ppdinteract} above). The 2:1 and 3:2 resonances are always stable. However, the 5:3 resonance goes unstable at very low eccentricities. Over half of the 5:3 resonances end up with a collision, resulting in a single planet with low eccentricity around 6 AU (recall that we stop integrations as soon as there is a collision or ejection). Since the planets have equal mass, about half of these cases eject the outer planet, and half the inner. The 5:3 resonance, however, was able to produce planets in a wide variety of resonances that included the 2:1, 3:2, 3:1, 5:3, 4:1, 6:1, and even a few cases of planets in a 10:1 and a possible 17:1 (where both planets held eccentricities above 0.95). These resonances were monitored by calculating the appropriate resonance variable defined by Equation \ref{eqn:resvar}. These latter resonances were produced by one of the planets being scattered outward due to instability.  With this planet now possessing a significant eccentricity, it was able to be captured into a higher-order resonance as it resumed inward migration. Both planets typically reached very high eccentricities during their coupled migration in these new resonances. Examples of a stable system and a scattered and recaptured system are given in Figures \ref{fig:21Stable} and \ref{fig:53Toss}.

	\subsubsection{Overall Results Without Eccentricity Damping}
	
We now repeat the simulations with eccentricity damping completely removed, allowing us to see how the outcomes look in the limit of maximal eccentricity excitation.  They are also presented in Table \ref{tab:resoutcome} for easy comparison.  

There is no significant change in the 2:1 and 5:3 resonances. The biggest change comes with the 3:2 resonance, which now rarely survives the migration towards the star. The majority of the 3:2 cases result in a collision. Also, ejections are more common for the 3:2 resonance than for the 5:3. The 3:2 also has a few cases where the planets are scattered and recaptured into another resonance, the possible resonances comprising a smaller subset of those mentioned above. 

	\subsubsection{The 5:3 Resonance}
\begin{figure}
\plotone{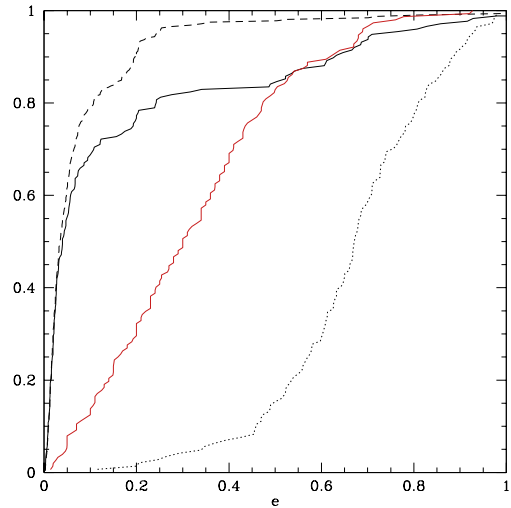}
\caption{Cumulative distribution of the remaining planet's eccentricity after a collision or ejection occurred due to an instability in the 5:3 resonance (black dashed and dotted-lines, respectively). The solid red line shows the distribution of observed extrasolar planetary systems. The selection method is the same as in Figure \ref{fig:ResCumHisto}. The black solid line is a randomly selected distribution including both collisions and ejections, which included the results of 25 integrations that resulted in an ejection and 150 integrations that resulted in a collision. The fraction taken from each was determined by their relative percentages of occurrence, which are given in Table \ref{tab:resoutcome}. } 
\label{fig:53CumHisto}
\end{figure}

The 5:3 resonance is worth discussing in more detail. For the case where eccentricity damping is present, we tally the distribution of eccentricities of the remaining planet after a collision or ejection has occurred. This are shown in Figure~\ref{fig:53CumHisto}. Additionally, Table \ref{tab:rescap} gives the various resonances (with relative capture probabilities) produced by instabilities with the 5:3 resonance. 

Collisions leave the remaining merged planet with very low eccentricities, although there is a larger range in their semi-major axes compared to our earlier integrations (See Figure \ref{fig:ResCumHisto}). Roughly 30\% of the planets are left closer than 5 AU, the minimum initial semi-major axis of the inner planet. The rest are found between 5 and 7 AU.

Ejections tend to leave the remaining planet at less than 3 AU with eccentricities above 0.6, and a large percentage of these planets having eccentricities between 0.6 and 0.8. This is similar to the results from \S \ref{sec:resonance stability}, where the eccentricity distribution approximately spanned the interval $[0.5,0.8]$. 

Similar to the procedure used in \S \ref{sec:resonance stability}, we also compare an overall distribution, but this time focusing entirely on the 5:3 resonance. Table \ref{tab:rescap} shows that in the case with eccentricity damping, the ratio of collisions to ejections is 6:1. Therefore, we randomly select 25 integrations resulting in an ejection, and 150 integrations resulting in a collision and plot the distribution in Figure \ref{fig:53CumHisto}. It is clear that the 5:3 resonance alone cannot accurately reproduce the observed eccentricity distribution in extrasolar planetary systems below eccentricities of 0.5.

\begin{deluxetable}{|c|l|c|c|}
\tablecaption{Relative percentages for the final possible outcomes given that two 1 $M_J$ planets are caught into a particular resonance and undergo disk-planet interactions. Cases with and without constant eccentricity damping are presented. A run is ``Stable" if the planets remain in that resonance throughout the entire integration, resulting in one of the planets migrating into the star. Cases where an ``Ejection" or ``Collision" occurred are noted appropriately. If the planets undergo instability and end up locking in another resonance for the majority of the integration, it is labeled as a ``Scattered and Recaptured."}
\tablewidth{0pt}
\startdata \hline
Resonance & Final Outcome & $t_e/t_m=0.1$ & $t_e=\infty$  \\  \hline
2:1 & Stable & 100\% & 100\% \\ 
 & Collision & 0\% & 0\% \\
 & Ejection & 0\% & 0\% \\
 & Scattered and Recaptured & 0\% & 0\% \\ \hline
3:2 & Stable & 100\% & 3\% \\
 & Collision  & 0\% & 68\% \\
 & Ejection & 0\% & 27\% \\
 & Scattered and Recaptured & 0\% & 2\% \\ \hline
5:3 & Stable & $<1$\% & 0\% \\
  & Collision & 54\% & 54\% \\
  & Ejection & 9\% & 12\% \\
  & Scattered and Recaptured & 36\% & 34\% \\ \hline     
\enddata
\label{tab:resoutcome}
\end{deluxetable}

\begin{deluxetable}{|c|c|c|}
\tablecaption{Possible resonances that two 1 $M_J$ planets were thrown into after they locked into a 5:3 resonance and ultimately went unstable due to eccentricity excitation and disk-planet interactions. The results are presented for the case where damping was present. Resonances are  included if coupled migration results after the two planets fall into these particular resonances such that the ratio of periods remained fixed for the migration. Additionally, we monitor that the corresponding resonance variables, defined by Equation \ref{eqn:resvar}, librate with amplitude less than $2\pi$ or undergo systematic periodic libration. See Figure \ref{fig:53Toss} for an example. The overall relative percentages of capture for each resonance is given.}
\tablewidth{240pt}
\startdata \hline
Order & Resonance & Relative \%  \\  \hline
First-Order 	& 2:1 & 21.3\% \\
			& 3:2 & 1.7\% \\ \hline
Second-Order	& 3:1 & 37.1\% \\ 
			& 5:3 & -- \\ \hline
Third-Order	& 4:1 & 4.9\% \\
			& 5:2 & 13.5\% \\ \hline
Other		& 5:1 & 1.1\% \\
			& 6:1 & 3.1\% \\
			& 7:2 & $<1\%$ \\
			& 7:1 & 5.1\% \\
			& 8:1 & 2.3\% \\
			& 9:1 & 2.8\% \\
			& 10:1 & 2.8\% \\
			& 11:1 & 1.7\% \\
			& 12:1 & 1.4\% \\
			& 14:1 & $<1\%$ \\
			& 17:1 & $<1\%$ \\ \hline
\enddata
\label{tab:rescap}
\end{deluxetable}

\section{Summary and Discussion}
In this paper, we have examined the stability of particular first and second-order resonances  with and without disk interactions. We have shown that the two-planet stability criterion (computed to third-order) is invalid for these resonances (see Figures \ref{fig:21Stab}, \ref{fig:32Stab}, and \ref{fig:53Stab}). The 2:1 resonance has regions of stability that extend up to mass ranges comparable to brown dwarfs, while the 3:2 resonance has a region of stability with combined planet masses up to twice the above criterion. The 5:3 resonance has a reduced region of stability compared to analytic criteria. 

Next, we have examined the results of instability. As shown in Figure \ref{fig:ResCumHisto}, combining an equal number of unstable outcomes originating in each of the above three resonances (with appropriate relative fractions of ejections and collisions) yields an eccentricity distribution which fits, at least, as well as previous studies of dynamical instability in multi-planet systems (in particular \citealt{chatterjeeetal07} for the three-planet case, and \citealt{jurictremaine07} for larger ensembles).  This is an intriguing result which raises the possibility of a link between these MMRs and exoplanet eccentricities.  However, significant caution is warranted in interpreting it.  Our ``outcomes" represent snapshots at the time of instability, with no attempt made to model the subsequent effect of the disk on the remaining single planet.  Also, the way in which we assemble resonances in \S \ref{sec:initial conditions} is rather artificial, at least in comparison to convergent migration of \S \ref{sec:ppdinteract}.

With the addition of migration in \S \ref{sec:ppdinteract}, plus the eccentricity damping expected from disk interactions, we find the 2:1 resonance to be the most stable, with 100\% of the runs remaining stable even without eccentricity damping (see Table \ref{tab:resoutcome}). Additionally, the 3:2 resonance remains completely stable with our adopted damping prescription. A second-order resonance, the 5:3, has a high capture efficiency but goes unstable at very low eccentricities due to close encounters.  This means that in order for planets to become locked into the 3:2 resonance, their primordial period separation must generally lie between 5:3 and 3:2.  In this study, we selected the initial semi-major axes of the planets to coincide with the location where it is believed planets form \citep[see, e.g.,][]{kokuboida02,thommesduncanlevison03}, rather than allowing the planets to migrate from further out in the disk. Under this assumption that planets form between 5 and 10 AU, if planets were able to form at any location with equal probability, the probability that a planet forms between the 3:2 and 5:3 resonances of a planet located at 5 AU is 

\begin{equation}
	\frac{\pi\left(\left[(5/3)^{2/3} - (3/2)^{2/3} \right]\cdot 5\right)^2}{\pi\cdot5^2} \simeq 0.009.
\end{equation}
That is, the probability of forming such a configuration is about 1\%. 

To date, we have not discovered any planets in a closer than 2:1 resonance.  The lack of 5:3 planets is readily accounted for by this study:  Since this resonance is unstable at low eccentricity, it is unstable to survive all the way to a mature planetary system.  The fact that 3:2 planets are {\it also} not seen, notwithstanding the resonance's robustness, may well be telling us something fundamental about how planets form:  It may simply be that neighboring giant planets are seldom born with primordial period separations of {\it less} than 5:3.  

The planets in HD 12661 are believed to be in a 6:1 resonance, and there is much debate as to how they could have become locked in such a high-order resonance. This study suggests a mechanism. We have shown that an instability in one of the lower-order resonances, which needs little to no initial eccentricity for planets to become trapped in it, can result in the planets scattering apart, to then be brought together again and re-locked in a more distant, higher-order resonance (See Table \ref{tab:rescap}). Further work is necessary to include the adjustment of the disk edge after the planets have been locked into a more distant MMR.  However, migration of the outward-scattered planet would likely be somewhat more gentle in a self-consistent disk---generally we would expect an annulus of gas to initially exist between the two planets---in which case resonant re-capture would actually tend to be {\it more} likely than in our current simple implementation.

Aside from the physical findings, this study highlights an important point on orders of accuracy. Since the resonance stability tests relied on two planets of unequal mass, we chose to use up to the third order term in Equation \ref{eqn:glad}, because this was the first term that was not symmetric in the masses (that is, a mapping $M_i \leftrightarrow M_o$ would not be an identity mapping).  Furthermore, as shown by Table 1, additional terms can change the result up to twice  the first-order value. If we were to truncate Equation \ref{eqn:glad} to just the first term, we would have concluded that the equation gives a better representation of the region of stability for the resonant case than it actually does.

This paper represents a first step in our study of resonances in protoplanetary discs. Here we have constructed systems where resonance encounters were impossible to avoid: requiring that we assumed the planets were coplanar and existed near these resonances without going unstable. A natural question this paper does not address is whether the dynamics that might occur due to relative inclinations affects the overall stability of the resonant systems. It has been shown \citep{thommeslissauer03} that it is possible for planets starting in a 2:1 resonance to later enter an inclination resonance, which generates large mutual inclinations even in an initially almost coplanar system.  It is not clear \textit{a priori}  whether this would promote dynamical instability or protect against it. Additionally, we have modeled the eccentricity damping time scale as a constant proportional to the migration time scale. Recent studies \citep{ogilvielubow03,moorheadadams07} have shown that damping may not be uniform and depends on the local properties of the disk as well as the overall geometry of the planets. In particular, $de/dt$ may even change sign depending on the value of the eccentricity. Therefore, it is important to better understand the nature of eccentricity growth and damping in order to understand the long-term stability of these resonances in the protoplanetary disk. 

Our improved distribution shown in Figure \ref{fig:ResCumHisto} was derived by using an equal number of cases from each of the three resonances studied. However, Table \ref{tab:resoutcome} suggests that we should find more planets in the 2:1 resonance compared to the 3:2 resonance during the disk's lifetime when we assume that all the planets migrated from a region farther out than the 2:1 resonance. In the case of gravitational instability, it may be assumed that planets could form in a more equally distributed manner before undergoing migration and locking into these resonances. Complete three dimensional integrations would allow us to gain an accurate representation of the distribution of resonance captures due to migration alone. Further stability tests will involve allowing one planets to migrate from greater distances, where near-resonance interactions cannot affect the overall stability of the system until the planets (given they remain stable) can migrate close enough together to encounter a particular resonance. 

Additionally, gap-forming planets on the edge of a disk in general ought to still be accreting some mass across their gap, and therefore mass growth will be taken into consideration. In this case, planets could lock into resonance under conditions we have shown to be stable, and through mass accretion enter a regime that analytic and numerical criteria determine to be unstable. A comparison of whether this region of stability is similar to what we have found here for constant-mass planets would give us further important insights into the possible range of planetary system configurations.  

We have gained a clearer picture of the range of possible dynamics within a protoplanetary disk. In doing so, we have seen how resonances can strongly affect planetary systems, long before the surrounding disk material has dissipated. This suggests that what happens before the disk disappears is key in determining a planetary system's ultimate dynamical architecture. One possible outcome is a system which, after the disk is gone, is left with planets in resonance.  In order for this to happen, the planets must enter these resonances relatively late in the disk's lifetime, so that they ultimately survive the coupled migration.  After the disk has dissipated, some resonances may eventually become unstable. Another possibility we have demonstrated is that resonances can be broken by dynamical instability that occurs while the disk is still present.  Our results allow us to make a connection between the planet formation process and observations of mature, resonant planetary systems, none of which have yet been observed in a closer than 2:1 resonance:  For planets that form further apart than a period ratio of 2:1 and migrate convergently, the very stable 2:1 constitutes a formidable barrier.  Between 2:1 and 5:3, the initial outcome is likely to be capture into 5:3, followed by a one-way trip to instability.  The 3:2 resonance is quite stable (unless eccentricity damping by the disk is absent), so provided that an appreciable number of planets form with a primordial period separation of less than the 5:3, we would expect to find some planets in a 3:2 resonance.  Thus, if future observations continue to reveal no such systems, then this may suggest a lower limit to orbital separations with which neighboring planets first form.

\acknowledgments
This work was supported by NSF Grant AST-0507727 at Northwestern University. ET would like to acknowledge the support of the Spitzer Theoretical Research Program, and NSERC of Canada.

\clearpage


\vfill
\begin{thebibliography}{}

\bibitem[Barnes \& Greenberg (2007)]{barnesgreen07} Barnes, R. \& Greenberg, R. 2007, \apjl, 665, L67

\bibitem[Bodenheimer et al. (2000)]{bodenheimeretal00} Bodenheimer, P., Hubickyj, O., \& Lissauer, J.~J. 2000, \icarus, 143, 2

\bibitem[Boss (2000)]{boss00} Boss, A.~P. 2000, \apj, 536, 101

\bibitem[Cameron (1978)]{cameron78} Cameron, A.~G.~W. 1978, Moon Planets, 18, 5

\bibitem[Chatterjee et al. (2007)]{chatterjeeetal07} Chatterjee, S., Ford, E.~B., \& Rasio, F.~A. 2007,  \textit{Accepted to \apj} (arXiv:astro-ph/0703166v2)

\bibitem[Fischer et al. (2003)]{fischeretal03} Fischer, D. Marcy, G.~W., Butler, R.~P., and Vogt, S.~S., Henry, G.~W., Pourbaix, D., Walp. B., Misch, A.~A., \& Wright, J.~T. 2003, \apj, 586, 1394

\bibitem[Ford et al. (2001)]{fordetal01} Ford, E.~B., Havlickova, M., \& Rasio, F.~A. 2001, \icarus, 150, 303

\bibitem[Ford \& Rasio (2007)]{rasioford07} Ford, E.~B., \& Rasio, F.~A. 2007, \textit{Accepted to \apj\ } (arXiv:astro-ph/0703163v2)

\bibitem[Gladman (1993)]{gladman93} Gladman, B. 1993, \icarus, 106, 247

\bibitem[Goldreich \& Tremaine (1979)]{goldtrem79} Goldreich, P., \& Tremaine, S. 1979, \apj, 233, 857

\bibitem[Goldreich \& Tremaine (1980)]{goldtrem80} Goldreich, P., \& Tremaine, S. 1980, \apj, 241, 425

\bibitem[Goldreich \& Sari (2003)]{goldreichsari03} Goldreich, P., \& Sari, R. 2003, \apj, 585, 1024

\bibitem[Juric \& Tremaine (2007)]{jurictremaine07} Juric, M., \& Tremaine, S. 2007, \textit{Accepted to \apj} (arXiv:astro-ph/0703160v2)

\bibitem[Kley et al. (2004)]{kleyetal04} Kley, W., Peitz, J., \& Bryden, G. 2004, \aap, 414, 735

\bibitem[Kokubo \& Ida (2002)]{kokuboida02} Kokubo, E., \& Ida, S. 2002, \apj, 581, 666

\bibitem[Lee \& Peale (2002)]{leepeale02} Lee, M.~H., \& Peale, S.~J. 2002, \apj, 567, 59

\bibitem[Lee et al. (2006)]{leeetal06} Lee, M.~H., Butler, R.~P., Fischer, D.~A., Marcy, G.~W., \& Vogt, S.~S. 2006, \apj, 641, 1178

\bibitem[Lin \& Papaloizou (1979)]{linpapa79} Lin, D.~N.~C., \& Papaloizou, J.~C.~B. 1979, \mnras, 186, 799

\bibitem[Lin \& Papaloizou (1986)]{linpapa86} Lin, D.~N.~C., \& Papaloizou, J.~C.~B. 1986, \apj, 309, 846

\bibitem[Lin \& Papaloizou (1993)]{linpapa93} Lin, D.~N.~C., \& Papaloizou, J.~C.~B. 1993, in \textit{Protostars and Planets III} ed. E.~H. Levy \& J.~I. Lunine (Tucson: University of Arizona Press) p. 749

\bibitem[Lissauer (1993)]{lissauer93} Lissauer, J.~J. 1993, \araa, 31, 129

\bibitem[Marcy \& Butler (1995)]{marcybutler95} Marcy, G.~W., \& Butler, R.~P. 1995, \textit{187th AAS Meeting BAAS}, 27, 1379

\bibitem[Marcy \& Butler (1998)]{marcybutler98} Marcy, G.~W., \& Butler, R.~P. 1998, \araa, 36, 57

\bibitem[Marcy et al. (2001)]{marcyetal01} Marcy, G.~W., Butler, R.~P., Fischer, D., Vogt, S.~S., Lissauer, J.~J., \& Rivera, E.~J. 2001, \apj, 556, 296

\bibitem[Mayor \& Queloz (1995)]{mayorqueloz95} Mayor, M., \& Queloz, D. 1995, \nat, 378, 355

\bibitem[Moorhead \& Adams (2007)]{moorheadadams07} Moorhead, A.~V., \& Adams, F.~C. 2007, \icarus, 708

\bibitem[Murray \& Dermott (2000)]{murraydermott} Murray, C.~D., \& Dermott, S.~F. 2000, Solar System Dynamics (Cambridge: Cambridge University Press)

\bibitem[Nelson et al. (2000)]{nelsonetal00} Nelson, R. P., Papaloizou, J.~C.~B., Masset, F., \& Kley, W. 2000, \mnras, 318, 18

\bibitem[Ogilvie \& Lubow (2003)]{ogilvielubow03} Ogilvie, G.~I., \& Lubow, S.~H. 2003, \apj, 587, 398

\bibitem[Papaloizou \& Lin (1984)]{papalin84} Papaloizou, J.~C.~B., \& Lin, D.~N.~C. 1984, \apj, 285, 818

\bibitem[Papaloizou \& Terquem (2001)]{papaterq01} Papaloizou, J~C.~B., \& Terquem, C. 2001, \mnras, 325, 221

\bibitem[Papaloizou et al. (2001)]{papaetal01} Papaloizou, J.~C.~B., Nelson, R.~P., \& Masset, F. 2001, \aap, 366, 263

\bibitem[Papaloizou (2003)]{papa03} Papaloizou, J.~C.~B. 2003, Celest. Mech. Dyn. Astro., 87, 53

\bibitem[Papaloizou et al. (2007)]{papaetal07} Papaloizou, J.~C.~B., Nelson, R.~P., Kley, W., Masset, F.~S., Artymowicz, P. 2007, in \textit{Protostars and Planets V} ed. B. Reipurth, D. Jewitt, \& K. Keil (Tucson, Univ. of Arizona Press)

\bibitem[Press et al. (2007)]{press07} Press, W.~H., Teukolsky, S.~A., Vetterling, W.~T., \& Flannery, B.~P. 1992, Numerical Recipes in C. The art of scientific computing (2nd ed.; Cambridge University Press)

\bibitem[Rasio \& Ford (1996)]{rasioford96} Rasio, F.~A., \& Ford, E.~B. 1996, Science, 274, 54

\bibitem[Snellgrove et al. (2001)]{snellgroveetal01} Snellgrove, M., Papaloizou, J.~C.~B., \& Nelson, R.~P. 2001, \aap,  374, 1092

\bibitem[Thommes \& Lissauer (2003)]{thommeslissauer03} Thommes, E.~W., \& Lissauer, J.~J. 2003, \apj, 597, 566

\bibitem[Thommes et al. (2003)]{thommesduncanlevison03} Thommes, E.~W., Duncan, M.~J., \& Levison, H.~F. 2003, \icarus, 161, 431

\bibitem[Thommes et al. (2008)]{thommesetal08} Thommes, E.~W., Matsumura, S., \& Rasio, F.~A. 2008, {\it Science}, 808

\bibitem[Trilling et al. (1998)]{trillingetal98} Trilling, D.~E., Bnez, W., Guillot, T., Lunine, J.~I., Hubbard, W. B., \& Burrows, A. 1998, \apj, 500, 902

\bibitem[Udry et al. (2007)]{udryetal07} Udry, S., Fischer, D., \& Queloz, D. 2007, in \textit{Protostar and Planets V} ed. B. Reipurth, D. Jewitt, \& K. Keil (Tucson, Univ. of Arizona Press)

\bibitem[Ward (1986)]{ward86} Ward, W.~R. 1986, \icarus, 67, 164

\bibitem[Ward (1997)]{ward97} Ward, W.~R. 1997, \icarus, 126, 261

\bibitem[Weidenschilling \& Marzari (1996)]{weidenschillingmarzari96} Weidenschilling, S.~J., \& Marzari, F. 1996, \mnras, 180, 57

\end{thebibliography}
\end{document}